# A Formal Transformation Method for Automated Fault Tree Generation from a UML Activity Model

C. E. Dickerson, *Senior Member, IEEE*, R. Roslan, and Siyuan Ji *Member, IEEE*

*Abstract*—Fault analysis and resolution of faults should be part of any end-to-end system development process. This paper is concerned with developing a formal transformation method that maps control flows modeled in UML Activities to semantically equivalent Fault Trees. The transformation method developed features the use of propositional calculus and probability theory. Fault Propagation Chains are introduced to facilitate the transformation method. An overarching metamodel comprised of transformations between models is developed and is applied to an understood Traffic Management System of Systems problem to demonstrate the approach. In this way, the relational structure of the system behavior model is reflected in the structure of the Fault Tree. The paper concludes with a discussion of limitations of the transformation method and proposes approaches to extend it to object flows, State Machines and functional allocations.

ACRONYMS

| | |
|---|---|
| AM-FPC-FT | Activity model-Fault Propagation Chain-Fault Tree |
| ARP | Aerospace Recommended Practice |
| CFT | Component Fault Tree |
| CMF | Conjunctive Material Form |
| CoO | Concept of Operation |
| FPC | Fault Propagation Chain |
| FTA | Fault Tree Analysis |
| FTM | Fault Tree Metamodel |
| MBSE | Model-based Systems Engineering |
| RAM | Reduced Activity Metamodel |
| RMS | Ramp Meter System |
| SoS | System of Systems |
| SoSE | System of Systems Engineering |
| SysML | System Modeling Language |
| UML | Unified Modeling Language |
| TCC | Traffic Control Center |
| TMSoS | Traffic Management System of Systems |

Manuscript submitted November 27[th], 2017. This work was supported in part by the Program for Simulation Innovation (PSi), a partnership between Jaguar Land Rover and UK EPSRC under Grant EP/K014226/1.
C. E. Dickerson, R. Roslan, and S. Ji (corresponding author) are with the Wolfson School of Mechanical, Electrical, and Manufacturing Engineering, Loughborough University, United Kingdom (e-mails: c.dickerson, r.roslan and s.ji@lboro.ac.uk).

## I. INTRODUCTION

Since the birth of Fault Tree Analysis (FTA) in 1961 at Bell Telephone Laboratories [1], FTA has become one of the most popular techniques for system safety assessment. More recently, from the 1990s to the 2000s, researchers and practitioners have been automating Fault Tree generation to reduce the time required by producing Fault Trees manually and to avoid human errors in manual Fault Tree constructions [2],[3]. The key challenge in the automation process, as identified by researchers in the field, is actually not the automation itself. Rather the challenge is to do with how the system should be modeled [2]. For instance, creating a comprehensive model for a complex system while maintaining sufficient level of detail can be challenging. In addition, a good modeling technique in one domain may not be general enough to have a wide applicability to cover other engineering domains. To tackle this challenge, modeling techniques developed for automated Fault Tree generation in the past include: model-based approaches such as the use of diagraphs [4],[5], state diagrams [6],[7], component-based modeling [2],[3], and knowledge-based approaches as reviewed in [8]. Despite the various level of success in the adoption of these methods in industrial settings, model availability has been one of the key issues. For instance, a substantial effort may be required to create the sophisticated system models, which is in contradiction to the idea of reducing time required in producing Fault Trees.

Much more recently, Model-based Systems Engineering (MBSE), a rapidly growing field originated from defense and aerospace, has attracted attention from the reliability and safety community [9],[10],[11]. In MBSE, modeling techniques and languages have been developed to model complex systems and System of Systems (SoS). Of the various modeling languages, Unified Modeling Language (UML) [12], which was originally developed for Software Engineering, became popular in Systems Engineering due to its general applicability and extendibility. UML has been elaborated into domain-specific languages such as the System Modeling Language (SysML) for general purpose system modeling [13] and the UML Profile for Modeling and Analysis of Real-time Embedded Systems for embedded systems [14]. Unlike the previous specific modeling techniques developed for automatic Fault Tree generation, UML and its extensions can facilitate the modeling of system across various domains. Nowadays, domains have started adopting MBSE approaches using UML; and its extensions are becoming widely adopted. Moreover, researchers and practitioners have also started using UML and its extensions for modeling and analysis from a system safety and reliability perspective [10],[11],[12],[13],[14],[15]. As such, for the purpose of automated Fault Tree generation, model availability is less an issue. However, efficient and reliable transformation techniques between models created using these languages and Fault Trees remain a challenge.

Thanks to the development of Transformation Languages, such as ATL Transformation Language developed by French Institute for Research in Computer Science and Automation [16] conforming to the Object Management Group standard, Query/View/Transformation [17], transformations between models developed in different languages and in different domains can be more easily achieved technically.

Researchers and practitioners have developed various ways for transforming UML models [18],[19],[20],[21] and SysML models [22],[23],[24] to Fault Trees. These transformations share the following commonalities: (i) they define entity-to-entity types of mapping. For instance, a Use Case in UML is mapped onto an intermediate event in a Fault Tree [20]; (ii) the transformations developed lack formality; hence they lack provable rationales to support the defined mapping.

Given that Fault Trees are meant to be used to assess system reliability and safety, trustworthiness of an automatically generated Fault Tree from system models can become questionable if the transformation does not have a formal basis. To bring formality into the development of system models and Fault Trees, attempts in formalizing UML models [11],[25] and Fault Trees models [26],[27] have been made independently. Further, to enhance consistencies between models developed in different domains, major projects such as Comprehensive Modeling for Advanced Systems of Systems [28] have developed tools and methods for engineers from all range of disciplines to provide support in cross-domain collaborations [29],[30]. Nonetheless, a formalized transformation method between UML models and Fault Tree models is still missing.

Therefore, in this paper, we seek to bridge this gap by developing a mathematically meaningful transformation method for generating Fault Trees from UML models. In this initial investigation, we focus mainly on system behavior in order to limit the scope of the research. To facilitate the automation of the developed method, e.g. to implement the transformation in ATL, we also abstract the transformation concept into an overarching metamodel that bridges a partial UML metamodel and a Fault Tree metamodel. Our main contributions are: (i) a transformation method comprised of a set of semantic mapping rules developed based on propositional calculus and probability theory to formally map Activity models into Fault Trees; (ii) the concept of Fault Propagation Chains as an intermediate step in the transformation method; and (iii) an overarching metamodel comprised of transformations between Activity models and Fault Tree models to facilitate automated Fault Tree generation. Unlike existing mapping techniques that focus on entity-to-entity mappings, our method reveals an important finding where the formal approach suggests that mappings should be based on the relational structure between entities.

To normalize the use of languages from different domains, in our work, we adopt definitions of terminologies used in the following two standards: UML Specification 2.5 [31] and ARP 4761 [32].

The rest of the paper is structured in the following way: in Sections II and III, background knowledge on model-based approaches to behavior modeling using UML Activities and to fault modeling using FTA are provided respectively. In addition, metamodels for the two domains are constructed based on existing standards. Then, Section IV presents the formalization of the model-based approach using propositional calculus. Based on the formalization and probability theory, we develop the mapping rules for transforming Activity model to Fault Trees in Section V. In addition, the transformational method is further abstracted into an overarching metamodel in this section. To evaluate the transformations developed, we apply them to a traffic management system case study in Section VI. Finally, in Section VII, we conclude our contributions and discuss the limitations of our method, then propose research directions for future work.

## II. MODEL-BASED APPROACH TO BEHAVIOR MODELING

The behaviors of a system directly associate to the execution of system functions. And the execution of a system function often leads to a change of system states. Hence, the modeling of the behaviors of a system can take a functional viewpoint or a state viewpoint. In model-based approach with UML, these two viewpoints lead to two different model representations: Activities and State Machines respectively. This paper focuses on Activity model for the purpose of demonstrating the development of an overarching metamodel. However, as discussed in the Conclusion, the extension of the metamodel to include State Machine Diagram is a necessary future step.

According to International Council on Systems Engineering (INCOSE) [33], behavior of a system can be classified into two types: dynamic and emergent. Dynamic behavior of a system is based on the time evolution of the system state, whereas emergent behavior can be seen collectively in a large scale as it cannot be understood in terms of individual system elements [33]. In this paper, dynamic behavior is the primary concern.

### A. Activity Diagram as a Behavior Model

UML Activities specify the sequencing of actions of a system or SoS. Here, *Action* is the technical term and a metaclass used in UML to represent the fundamental unit of behavior specification [31]. For the rest of the paper, for clarity, we adopt Pascal case format, i.e. concatenating capitalized words, in the naming of metaclasses, e.g. ActivityNode. In the case of modeling system behavior with Activities, an action, represented by a node (ExecutableNode), refers to an elementary step to be executed by the system. To show the sequencing of the steps, the nodes are then connected via edges (ActivityEdge). Table I provides a list of graphical notations available to the modeling of system behavior in Activities within the scope of this paper. In addition to an executable node, Activity modeling also involves the use of other types of nodes, e.g. control nodes, and objects. As objects do not represent system functions, they will not be considered in this paper for the purpose of functional fault analysis. This also reduces the graphical complexity of an Activity model to only include control flows.

TABLE I
ACTIVITY MODEL SYMBOLS

| Symbol | Description |
| --- | --- |
| | **Action** – action state of system behavior |
| | **Control Flow** – directional activity flow of control nodes and action states |
| | **Initial Node** – initial state of activity flow |
| | **Activity Final Node** – final state of activities completion |
| | **Decision Node** – point of alternate paths decision |
| | **Merge Node** – point of multiple flows merge to a single flow without synchronization |
| | **Fork Node** – point of single flow splits to multiple flows |
| | **Join Node** – point of multiple flows synchronize to a single flow |
| | **Swimlane** – classify and hold activity flows according to systems in partitions |

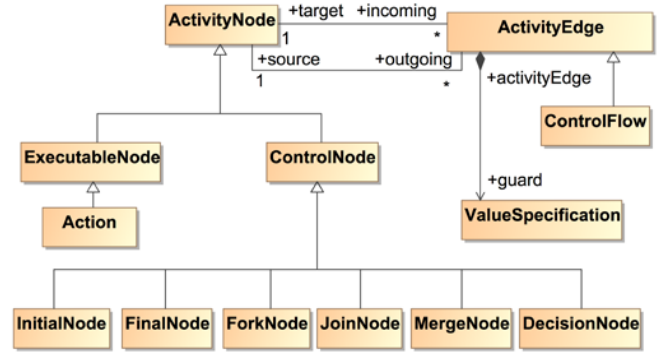

Fig. 1. Reduced Activity Metamodel extracted from the UML Specification [31].

To describe the execution of an action and the flow of controls, UML uses the concepts of tokens (which are not explicitly modeled in Activities) and guards. After a function completes its execution, a control token will be offered to the next node via the edge that connects them. Furthermore, edges may have guards on them. A token can only pass through an edge with a guard if the guard evaluates the token to true for the offered token. Guard is commonly used on the outgoing edges of a decision node (DecisionNode). The concept of passing a token after the successful execution of an action provides the foundation for formalizing control flows by using propositional calculus.

*B. Activities Metamodel*

Activities Metamodel serves as a reference to Activity modeling by defining the abstract syntax and the interrelationship between model elements in the standard Activity model [34]. As depicted in Fig. 1, within the scope of this paper, only a subset of the UML Activities metamodel is considered for the development of the overarching metamodel. Metaclasses, such as Object and ObjectFlow are neglected due to their irrelevance to fault modeling from a functional viewpoint. In addition to this subset, Action metaclass is also included based on previous discussions. This metamodel will be referred to a Reduced Activity Metamodel (RAM) in the rest of the paper.

In the RAM, at the top level, ActivityNode and ActivityEdge are associated with each other through two relations. These relations represent the two cases in the modeling of control flow: (i) an edge comes after a node, and (ii) a node comes after an edge. Detailed descriptions of the lower level metaclasses are provided as follows:

1. The ActivityNode generalizes two types of nodes: ExecutableNode and ControlNode. These activity nodes are points of intersection where respective operation takes place in the Activity model.

2. The ExecutableNode is the generalization of Action which specifies the actions to be executed in the Activity model. The proposition that describes an action is captured within the Action symbol (Table I).

3. The ControlNode generalizes a set of paired nodes that are used to manage different types of control flows. These node pairs are: (i) InitialNode and FinalNode which are used to indicate the starting and ending point of a flow respectively; (ii) ForkNode and JoinNode which are used to specify concurrent flows; and (iii) DecisionNode and MergeNode which are used to specify alternative flows. Although the pairings are not reflected in the RAM, paired usage has been regarded as necessary practice for semantic consistency. As many of these control nodes allow multiple coexisting (concurrent and alternative) flows to be modeled, it is therefore possible to have situations where a single activity node is associated to multiple activity edges. These situations are covered by multiplicity on the association line where an asterisk symbol is depicted toward the ActivityEdge end.

4. In the UML 2.5 specification, guard is not explicitly captured by a metaclass [31]. Instead, the metamodel uses the

TABLE II
FAULT TREE SYMBOLS

| Symbol | Description |
|---|---|
| ▭ | **Output Event** – event which is an output of a logic symbol (refer as top event or intermediate event) |
| ○ | **Basic Event** – event which is internal to a system without further development |
| ⌂ | **External Event** – event which is external to a system |
| ◇ | **Undeveloped Event** – event which has little impact to the top event without further development |
| ⬭ | **Conditional Event** – event of a necessary condition for a failure mode to occur |
| △ | **Transfer Event** – event which indicates a fault tree information is transferred out to another fault tree or transferred into fault tree |
| ⌒ | **AND-Gate** – Boolean logic gate – output event occur when all intermediate events occurred |
| ⌒ | **OR-Gate** - Boolean logic gate – output event occur when at least one intermediate event occurred |
| ⌒ | **Priority AND-Gate** - Boolean logic gate – output event occur when all intermediate events occurred in a specific sequence (sequence usually represented by a conditional event ) |
| ⬡ | **Inhibit-Gate** – Boolean logic gate – output event occur with the presence of an enabling conditional event |

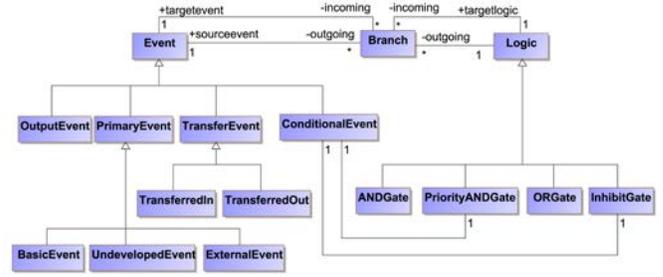

Fig. 2. Fault Tree Metamodel developed based on ARP 4761 [32].

concept of ValueSpecification.

### III. MODEL-BASED APPROACH TO FAULT ANALYSIS

Model-based approach to fault analysis, such as FTA, Reliability Block Diagrams, Binary Decision Diagrams, Dependency Diagram, and Markov Analysis are widely used for system safety assessment. Safety Analysts often use these model-based techniques to evaluate system architecture to quantify probabilities of occurrence of system failures, and to identify potential related risks. In this section, we provide essential background knowledge in FTA and construct a Fault Tree Metamodel (FTM) based on safety standard ARP 4761 [32].

*A. Fault Tree Analysis*

FTA is a conventional system analysis technique for assessing safety in the development of a system. It is a deductive failure analysis method for determining the causes of undesired events. Table II presents a list of basic notations in FTA used in this paper based on ARP 4761 [32]. IEC 61025, a specific standard of FTA, defines two types of FTA: static and dynamic [35]. Other methods such as dynamic FTA [36] and Fuzzy FTA [37] will not be considered in this paper. In brief, in conventional FTA, qualitative evaluation is concerned with identifying information such as failure paths with the use of minimal cut sets; and quantitative evaluation is concerned with determining the probability of the top event.

*B. Fault Tree Metamodel*

A standardized metamodel for Fault Tree currently remains unspecified [20]. Several research efforts have contributed to the construction of a generic metamodel for Fault Tree. For instance, the generic Component Fault Tree (CFT) Metamodel [38] and a Fault Tree metamodel [20] are developed based on different viewpoints. CFT Metamodel is constructed based on the hierarchical decomposition of a system. It emphasizes on the concepts of Component and Component Proxies based on component-based software development. Whilst, the latter metamodel is constructed based on the FaultCat analysis tool. Despite the different viewpoints, the two metamodels share a common set of metaclasses, e.g. Event and Gate (logic), which represent the backbone structure of a Fault Tree.

Metamodels of Fault Tree developed in the researches are further extended and applied in other areas. For instance, the CFT Metamodel has been further integrated with Architecture Domain Specific Modeling Language to reduce efforts and times needed for safety analysis; and the integration also leads to potential reusable models. The metamodel developed in [20] is being used for dependability analysis between UML model and Fault Tree. In this paper, metamodel of Fault Tree is developed for the purpose of unifying system functional architecture and system failure analysis.

The FTM developed in this paper, as depicted in Fig. 2, is constructed based on ARP 4761 [32]. Similar to the construction of the RAM, we use Pascal case format to denote a metaclass name to distinguish it from the name of an actual model element. Individual metaclass is explained as follows:

1. The main elements in Fault Tree model are events, branches, and logic gates. Hence, they are abstracted into metaclasses at the highest-level as Event, Branch, and Logic metaclasses respectively in the FTM. Similar to how ActivityEdge and ActivityNode are connected in the RAM, Branch is associated to both Event and Logic also by two-way relations. The multiplicity also reflects situations where an event (or logic gate) can be associated with one or more multiple branches going into the event (or logic gate) and out of the event (or logic gate).

2. Event metaclass consists of OutputEvent, PrimaryEvent, TransferEvent, and ConditionalEvent, each representing the corresponding type of event seen in Fault Trees (c.f. Table II). The PrimaryEvent further generalizes BasicEvent, UndevelopedEvent, and ExternalEvent. The TransferEvent is a generalization of TransferredIn and TransferredOut which represents the existence of external branch of Fault Trees.

3. Logic metaclass consists of ANDGate, PriorityANDGate, ORGate, and InhibitGate which is used to evaluate input branches and tie the branches together.

4. ConditionalEvent is associated with PriorityANDGate or InhibitGate. This reflects the use of conditional event in Fault Tree construction where a priority AND-gate and inhibit-gate is always accompanied with a conditional event specified in an oval shape (c.f. Table II).

## IV. FORMALIZATION USING PROPOSITIONAL CALCULUS

The model-based approach in Section II led to the specification of a UML metamodel for behavior modeling. These were adapted from currently accepted graphical models which are not formal. Formalizing these models in the propositional calculus will facilitate transformation between behavior and fault models.

### A. The Propositional Calculus

Proposition is an assertion that expresses premise conclusion or argument which can be expressed in symbol or variable [39]. The propositional calculus is the part of mathematical logic where the validity of an argument depends only on how the propositional sentences are formed and not on the internal structure of the propositions. This is sufficient for the modeling and analysis of functional faults. Specifically, the propositions of interest will be of the following form:

$$p_i: \text{The action } A_i \text{ completed execution.} \quad (1)$$

Note that this is a decidable declarative statement which has a yes-no answer; and therefore adheres Boolean calculus of evaluation of truth. The action $A_i$ can then be associated with {1,0} or {T, F} values. The truth values of the propositions will be regarded as outcomes of a designed experiment.

It is useful to view a coin tossing experiment from this behavioral viewpoint. The action $A_i$ can be stated as: the coin was flipped. Specifying the outcome is part of the design of the experiment. This could be as simple as specifying the coin began in one state {Head,Tail}, underwent a random change of state, and resulted in the coin coming to rest in a state determined by the side of the coin facing up when the process completed execution. This process could fail to complete execution if, for example, either change was not random or the coin somehow came to rest on its edge. The outcome of this behavior is silent on the end state of the coin. A second proposition could be introduced to complete the experiment; but this would not be a behavioral proposition.

It is important to note in this simple example, how the specification of the experiment immediately led to details associated with the internal structure of the behavioral proposition, $p_i$. Further analysis of behavior and faults involving nonfunctional properties will need the predicate calculus, which is a separate concern than what is presented in this paper.

### B. Representation of Behavioral Faults

The sentences of interest will be well-formed formulae consisting of the types of propositions in (1). These will be referred to as behavioral propositions. For any two propositions, $p_i$ and $p_j$, the sentences are declarations in one of the following forms or a negation of the form:

$$\neg p_i \quad p_i \wedge p_j \quad p_i \vee p_j \quad p_i \rightarrow p_j . \quad (2)$$

where $\neg p_i$ is the negation of $p_i$, $p_i \wedge p_j$ is a logical conjunction between two behavioral propositions; $p_i \vee p_j$ is a logical disjunction between two behavioral propositions; and $p_i \rightarrow p_j$ is material implication, which is read as $p_i$ implies $p_j$. The proposition $p_s$ will also be used and reserved for the completion of the overall execution of the system functions within a given Activity model.

The analyst has choices as to how faults should be presented. From a logical viewpoint it can be useful to consider collections of propositions in disjunctive normal form [40], for example:

$$(p_1 \wedge p_2) \vee (p_3 \wedge p_4) . \quad (3)$$

These might be associated with a logical transition to a system behavior $p_i$ or its failure $\neg p_i$. For the rest of paper, the negated proposition is also defined as a fault event,

$$a_i \stackrel{\text{def}}{=} \neg p_i: \text{The action } A_i \text{ failed to complete execution.} \quad (4)$$

Correspondingly, the proposition, $a_s$, is reserved for overall system failure. In the language of Fault Tree, this will be regarded as the top event within the scope of this paper.

### C. Control Flows as Implication Chains

A control flow models a sequence of actions in an Activity model. Two adjacent actions, $A_i$ and $A_{i+1}$ in a control flow can then be described as $A_i$ precedes $A_{i+1}$. Assuming that there are no other actions preceding $A_{i+1}$, one can infer that if $A_{i+1}$ has completed its execution, then the action $A_i$ must have also completed its execution. Using the propositions introduced in (1), we express the above statement logically by a material implication,

$$p_{i+1} \rightarrow p_i . \quad (5)$$

Control flows involving concurrent and alternative flows, can then be expressed by conjunctions and disjunctions of these implications to form a Logical Model, e.g. the control flows depicted in the Activity model in Fig. 8 of Section VI. In the following subsection, we will show how to transform these implications to describe fault structures.

### D. Fault Propagation Chains

To transform a Logical Model into behavioral fault representation, instead of using the negation of individual action proposition, e.g. $\neg p_i$, we apply the contrapositive to the material implication in (5). Incorporating the definition of a fault event given in (4), we define the contrapositive form of $p_{i+1} \to p_i$ as,

$$a_i \to a_{i+1} \overset{\text{def}}{=} \neg\, p_i \to \neg\, p_{i+1}\,. \tag{6}$$

This expression can be interpreted as follows: if the action $A_i$ failed to complete execution, then action $A_{i+1}$ will fail to execute. This statement also makes sense from the behavioral viewpoint described in an Activity model. For instance, if $A_i$ failed to complete execution, a token will not be generated and passed onto the next action, $A_{i+1}$. Hence, $A_{i+1}$ will not execute. The consequence of $A_{i+1}$ failing to execute can be therefore understood by the concept of fault propagation.

Using the contrapositive form in (6) as the basis, we further introduce a chain of material implications to model fault propagations in a control flow:

$$(a_1 \to a_2) \wedge (a_2 \to a_3) \wedge \ldots \wedge (a_{n-1} \to a_n)\,, \tag{7}$$

where each $(\cdots)$ describes a unit of fault propagation between two adjacent actions and a chain is only permitted to be a conjunction of these fault propagation units as defined in (6). Multiple chains may be combined by a conjunction or disjunction; and this will be discussed in Section V in detail. Some event structures may involve the use of disjunctions, for example:

$$a_i \vee a_j \to a_k\,. \tag{8}$$

By refactoring, a form using conjunctions instead, may be produced:

$$(a_i \to a_k) \wedge (a_j \to a_k)\,. \tag{9}$$

However, note that this is not a chain because the antecedent of the second material implication is not the consequent of the first material implication.

For situations where an event structure reads,

$$a_i \wedge a_j \to a_k\,, \tag{10}$$

we define the conjunction of the multiple fault events as a contracted fault event, for example, for two events in conjunction, we have

$$a_{i,j} \overset{\text{def}}{=} a_i \wedge a_j\,. \tag{11}$$

We define the logical forms used in (7), (9), (10) and (11) for describing fault event structures *Conjunctive Material Form* (CMF). Specifically, this is a conjunction of material implications in which the two propositions within each material implication can be a conjunction. The primary reason for introducing CMF and contracted fault events is to simplify the representation of a long logic formula into simpler chains of implications. Although not all logic formulae can be written in CMF, we will demonstrate in the next section that the transformation method is applicable to chains that are in CMF as well as combined chains that are not in CMF. In the rest of the paper, the graphical representation of chains of implications will be referred to as a Fault Propagation Chain (FPC).

## V. FROM ACTIVITIES TO FAULT TREES

In this section the logical formalization of UML Activities is used as the basis to develop a set of semantic mapping rules to transform system behavior as represented by Activity models to behavioral faults as represented by FPCs. Probability theory [41] is then applied to further transform the FPCs into a conventional Fault Tree representation. The section begins with transformation of elementary control flows modeled in UML Activity diagrams to semantically equivalent Fault Trees. Section VI will demonstrate the approach by applying the transformation method to an elementary case study that is well understood.

The application of the method to broader classes of engineering problems can be understood in two ways. The first includes complicated problems (i.e. with similar form but greater detail than the elementary forms and case study) that can be reduced to conjunctive structures. These will simply result in longer CMF forms, such as in (7), or longer conjunctions than the elementary form in (11). Complexity, on the other hand, can arise from combinations of disjunctive forms with conjunctive and material forms that do not necessarily admit reduction into a single chain. Complexity is addressed at the end of this section by means of transforming a nested control structure into a Fault Tree.

### A. Semantic Mapping Rules: Activities to Fault Propagation Chains

Based on the formalization introduced in the previous section, representative types of elementary control flows seen in typical Activity model are semantically mapped into logical models, i.e. implication chains, and then transformed into FPCs by applying the contraposition of the logical models. Finally, graphical representations of the FPC are depicted. Starting with the graphical control flows and ending with the graphical FPC, for every elementary control flow structure, each stage of the semantic mapping is depicted in Table III. Detailed discussion on the various control flow structures and their associated semantic mapping rules are provided below:

1. The control flow presented in (a) consists of two actions sequenced in series. As already discussed in Section IV, this control flow can be modeled by using propositional logic as $p_2 \to p_1$, i.e. the completed execution of $A_2$ implies that the execution of $A_1$ must have been completed. Then, applying contraposition to the logic model, we derive a fault propagation chain that reads $a_1 \to a_2$.

2. Row (b) and (c) are concerned with the control flows that involve a pair of folk and join nodes. In (b), one action flows into two concurrent actions via a fork node while in (c), two concurrent actions flow into a single action via a join node.

TABLE III
SEMANTIC MAPPING RULES

| Control Flow | Logical Model | Contrapositive | | | | Fault Propagation |
|---|---|---|---|---|---|---|
| 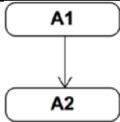 | $p_2$<br>$\downarrow$<br>$p_1$ | $\neg p_1$<br>$\downarrow$<br>$\neg p_2$ | $\overset{\text{def}}{=}$ | $a_1$<br>$\downarrow$<br>$a_2$ | | 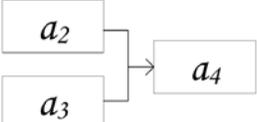 |
| (a) | | | | | | |
| 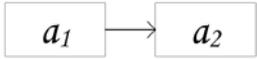 | $p_2 \lor p_3$<br>$\downarrow$<br>$p_1$ | $\neg p_1$<br>$\downarrow$<br>$\neg p_2$ $\land$ | $\neg p_1$<br>$\downarrow$<br>$\neg p_3$ $\overset{\text{def}}{=}$ | $a_1$<br>$\downarrow$<br>$a_2$ $\land$ | $a_1$<br>$\downarrow$<br>$a_3$ | 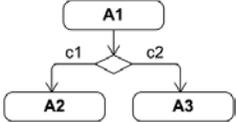 |
| (b) | | | | | | |
| 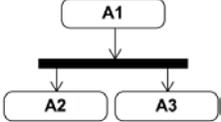 | $p_4$<br>$\downarrow$<br>$p_2 \land p_3$ | $\neg p_2$<br>$\downarrow$<br>$\neg p_4$ $\land$ | $\neg p_3$<br>$\downarrow$<br>$\neg p_4$ $\overset{\text{def}}{=}$ | $a_2$<br>$\downarrow$<br>$a_4$ $\land$ | $a_3$<br>$\downarrow$<br>$a_4$ | 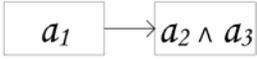 |
| (c) | | | | | | |
| 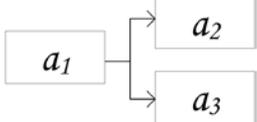 | $p_2 \lor p_3$<br>$\downarrow$<br>$p_1$ | $\neg p_1$<br>$\downarrow$<br>$\neg p_2 \land \neg p_3$ | $\overset{\text{def}}{=}$ | $a_1$<br>$\downarrow$<br>$a_2 \land a_3$ | | 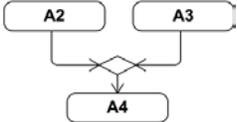 |
| (d) | | | | | | |
| 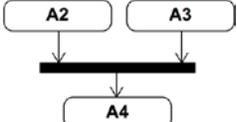 | $p_4$<br>$\downarrow$<br>$p_2 \lor p_3$ | $\neg p_2 \land \neg p_3$<br>$\downarrow$<br>$\neg p_4$ | $\overset{\text{def}}{=}$ | $a_2 \land a_3$<br>$\downarrow$<br>$a_4$ | | 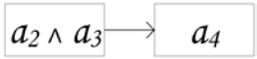 |
| (e) | | | | | | |

The logical model for (b) can be formulated as $p_2 \lor p_3 \to p_1$, which means that the completed execution of either $A_2$ or $A_3$ implies that $A_1$ has been executed. Applying contraposition to the logical model leads to the expression $a_1 \to (a_2 \land a_3)$ or equivalently, $(a_1 \to a_2) \land (a_1 \to a_3)$.

The logical model for (c) can be formulated as $p_4 \to (p_2 \land p_3)$, which means that the completed execution of $A_4$ implies that both $A_2$ and $A_3$ must have been completed. Applying contrapositive to the logical model leads to the expression $(a_2 \lor a_3) \to a_4$. However, this is not in the CMF introduced in Section IV. Hence, by using (9), we refactor the expression into a CMF that reads $(a_2 \to a_4) \land (a_3 \to a_4)$. Immediately, one realizes that how this CMF connects naturally with the previous CMF $(a_1 \to a_2) \land (a_1 \to a_3)$ in (b) to form two concurrent chains, $(a_1 \to a_2) \land (a_2 \to a_4)$ and $(a_1 \to a_3) \land (a_3 \to a_4)$. The FPC is depicted in the last column in (b) and (c), where $a_1$ bifurcates into two chains and converge back into one chain at $a_4$.

3. Row (d) and (e) are concerned with the control flows involving a pair of decision and merge nodes. Different from (b), one action can only flows into one of the two paths based on the decision being made. Hence, only one of the two later actions, $A_2$ and $A_3$ can be executed. Then, in (e), the two paths merge into a single path where the control sequence flows either from $A_2$ to $A_4$ or from $A_3$ to $A_4$.

Although the control flow in (d) is very different from (b), they share the same logical model $(p_2 \lor p_3) \to p_1$. This is because that no matter which path is taken, as long as there is a completed execution of either $A_2$ or $A_3$, $A_1$ must have been executed. Similarly, the contrapositive of the logical model can be expressed as $a_1 \to (a_2 \land a_3)$. Further using the definition of a contracted fault event introduced in (11), we arrive at a FPC for (d) that reads $a_1 \to a_{2,3}$.

The logical model for (e) can be formulated as $p_4 \to (p_2 \lor p_3)$ (this is different from (c)), which means that the completed execution of $A_4$ implies that either $A_2$ or $A_3$ has been executed. Applying contraposition to the logical model leads to the expression $(a_2 \land a_3) \to a_4$. Similarly, this expression becomes $a_{2,3} \to a_4$ by defining the conjunction on the left as a contracted fault event. The two FPC in (d) and (e) also connect naturally to form a single chain, $(a_1 \to a_{2,3}) \land (a_{2,3} \to a_4)$, where $a_{2,3} \overset{\text{def}}{=} a_2 \land a_3$ is a contracted fault event as defined in (11).

### B. Fault Propagation Chains: Probability Analysis

In conventional FTA, the edges connecting lower and higher level events do not possess a well-defined semantic meaning. For instance, an edge may mean composition where a higher level event is composed by a lower event and another different lower event by an AND-gate; an edge can also mean causality where an intermediate event is caused by either one of the basic events that are connected by an OR-gate. In contrast, the directed edges in the FPC proposed in this paper mean exactly material implications and fault propagates along these directed edges.

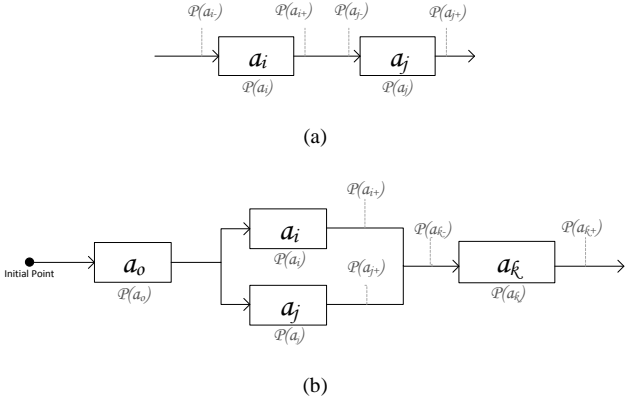

Fig. 3. Examples of Fault Propagation Chains: (a) series propagation; (b) bifurcated and merged propagation.

In the case of series (and non-exclusive), an FPC can be thought as the conjunction of a combination of non-repeating elementary units of the form $a_i \to a_j, (i \neq j)$ where $a_i$ can be a contracted fault event; and a disjunction in the case of exclusive chains. For every elementary unit, one can ask the question: what is the truth value of $a_j$ given a truth value of $a_i$ being True (or False)? The answer can be obtained from building a truth table: if $a_i$ is True, then $a_j$ is True; but if $a_i$ is False, $a_j$ can be either True or False. Assuming that one can define a probability for the fault event $a_i$ being True, we can then conduct an observation experiment to count the number of times that event $a_j$ has returned a True value. One would then observe that the probability of $a_j$ denoted as $P(a_j)$, is greater than the probability of $a_i$ denoted as $P(a_i)$, i.e. $P(a_j) \geq P(a_i)$. This mathematical fact, interpreted in the language of fault analysis, immediately leads to the concept of fault propagation. For instance, if the fault event, $a_i$ has happened, one conclude automatically that $a_j$ will happen; and if $a_i$ has not happened, $a_j$ may still happen.

To better describe the probability of the occurrence of these fault events, with the aid of Fig. 3(a), the following definitions and notations will be adopted:

Definition 1: $P(a_i)$ is defined as the probability of the fault event, $a_i$, independent of the occurrence of any other fault events, i.e. prior faults propagated to $a_i$, if any, are not counted toward $P(a_i)$.

Definition 2: $P(a_{i+})$ is defined as the probability of observing a system failure right after the (supposed) execution of $A_i$. Similarly, one can also define a probability measurement, $P(a_{i-})$, as the probability of observing a system failure just before the execution of action, $A_i$. In the simple construct shown in Fig. 3(a), one notices that $P(a_{i+}) = P(a_{j-})$. Note that $a_{i+}$ and $a_{i-}$ are fictitious events that are not originally presented in the Activity model.

Using the above definitions, for any given chain, $a_i \to a_j$, one can establish the following mathematical relations:

$$\begin{cases} P(a_{i+}) = P(a_{i-}) + P(a_i) - P(a_{i-})P(a_i) \\ P(a_{j+}) = P(a_{i+}) + P(a_j) - P(a_{i+})P(a_j) \end{cases}, \quad (12)$$

with $P(a_{i+}) = P(a_{j-})$. To see how these relations are derived, let us consider the following thought experiment [41]. An observer runs the system for $n$ times and observes the number of system failures, i.e. failed to complete the intended system behavior as modeled in the UML Activity. Up to the point at where system function $A_i$ is about to execute, the observer would measure a probability of system failure equal to $P(a_{i-})$ based on Definition 2. Then, statistically, there are $nP(a_{i-})$ number of experiments in which the system would fail at this point and $n(1 - P(a_{i-}))$ number of experiments in which the system successfully proceed to the execution of $A_i$. Given that there is a probability of failure in the execution of $A_i$, i.e. $P(a_i)$, based on Definition 1, in addition to the experiments in which the system failed before the execution of $A_i$, the observer would further expect a $m = n(1 - P(a_{i-}))(P(a_i))$ number of experiments in which the system would fail right after the supposed execution of $A_i$. Therefore, one can calculate $P(a_{i+})$ by the quotient of the total number of failed events at this point and the total number of experiments as,

$$\begin{aligned} P(a_{i+}) &= \frac{nP(a_{i-}) + m}{n} \\ &= P(a_{i-}) + (1 - P(a_{i-}))(P(a_i)) \\ &= P(a_{i-}) + P(a_i) - P(a_{i-})P(a_i). \end{aligned} \quad (13)$$

The second relation in (12) can be derived using exactly the same concept.

In a simplified case where $a_i$ is assumed to be the first fault event in the entire chain, i.e. $P(a_{i-}) = 0$, the relations reduce to:

$$P(a_{j+}) = P(a_i) + P(a_j) - P(a_i)P(a_j). \quad (14)$$

The relation in (13) and (14) will form the basis for the analysis of FPCs and the transformation of FPCs to Fault Trees.

As shown in Table III, certain control flows after the transformation can lead to bifurcation and convergence in the FPC. In these cases, the probability calculation needs to be carefully dealt with to avoid over-counting of the probability of repeated events. Using the example depicted in Fig. 3(b), based on the derived relations in (12) and (14), one can establish the following relations:

$$\begin{cases} P(a_{i+}) = P(a_0) + P(a_i) - P(a_0)P(a_i) \\ P(a_{j+}) = P(a_0) + P(a_j) - P(a_0)P(a_j) \\ P(a_{k-}) = P(a_{i+}) + P(a_{j+}) - P(a_{i+})P(a_{j+}|a_{i+}) \\ P(a_{k+}) = P(a_{k-}) + P(a_k) - P(a_{k-})P(a_k) \end{cases}, \quad (15)$$

with $a_0$ being an initial fault event, i.e. no other fault propagates to it. The first and second relations in (15) are derived based on the result in (14) by considering the bifurcation as two parallel (non-exclusive) chains, i.e. $a_0$ to $a_i$; and $a_0$ to $a_j$ respectively. The fourth relation can be derived using the same thought experiment in deriving the relations in (12).

The derivation of the third relation in (15) requires a consideration of the repeated event, $a_0$ in the convergence of the two paths. Instead of directly using the relations derived in (12), let us come back to the thought experiment in which the observer runs the system of $n$ times. Based on previous derivations, the observer would first expect $nP(a_0)$ number of system failures due to the failed execution of $A_0$. The observer would also expect an additional $m = n(1 - P(a_0))(P(a_i))$ number of system failures right after the supposed execution of $A_i$; and an additional $q = n(1 - P(a_0))(P(a_j))$ number of systems failures right after the supposed execution of $A_j$. As such, the total number of system failures right before the execution of $A_k$ is given by $nP(a_0) + m + q$. This then allows us to calculate $P(a_{k-})$, as

$$\begin{aligned} P(a_{k-}) &= \frac{nP(a_0) + m + q}{n} \\ &= P(a_0) + P(a_i) - P(a_0)P(a_i) + P(a_j) - P(a_0)P(a_j) \\ &= P(a_{i+}) + P(a_{j+}) - P(a_0), \end{aligned} \quad (16)$$

where the last step has used the first two relations in (15) as substitutions. In this representation, as seen in Fig. 3(b), $a_0$ is a repeated event, and hence, can be considered as the intersection of the two events, $a_{i+}$ and $a_{j+}$ in which

$$P(a_{i+} \cap a_{j+}) = P(a_{i+})P(a_{j+}|a_{i+}) = P(a_0), \quad (17)$$

where $P(a_{j+}|a_{i+})$ is the conditional probability of $a_{j+}$ given $a_{i+}$. Substituting (17) into (16) would arrive the third relation in (15). It is worth noting that due to the presence of the repeated event, $a_{i+}$ and $a_{j+}$ can be dependent. For instance, if an observation of $a_{i+}$ returns a True value due to the occurrence of the repeated event, $a_0$, immediately, one would expect $a_{j+}$ to return a True value as well. However, if the observation of $a_{i+}$ returns a True value due to the occurrence of $a_i$ instead of $a_0$, an observation of $a_{j+}$ may or may not return a True value depending whether $a_j$ has occurred or not. The situation where a repeated fault event happens in a FPC is analogous to a repeated basic event seen in a conventional Fault Tree.

A contracted fault event, such as $a_{i,j}$, can be viewed identically as a normal fault event such that it follows exactly the same analysis rule. However, from both the system modeling viewpoint and fault analysis viewpoint, it can be useful to understand how individual fault event embedded in the contraction can affect the overall FPC. Tracing back to the Activity model, a contracted event embeds exclusive paths, i.e. if one path is taken, no other paths will be taken. Theoretically, it is possible to attribute each of the paths a probability for which it might be taken. The sum of the probabilities has to be exactly one; this is to reflect that a path is always taken. Now, coming back to the fault analysis viewpoint, an embedded fault event then becomes only relevant when its corresponding path (in the Activity model) is taken. This allows one to establish the mathematical relation as follows:

$$P(a_{i,j}) = P(a_i)P(c_i) + P(a_j)P(c_j), \quad (18)$$

where $P(c_i)$ is the probability of the system taking the path, $c_i$. Note also that $c_i$ and $c_j$ are not fault events and that they are mutually exclusive. The conditional statement, $c_i$, traces back to the guard contents on an activity edge in a control flow that contains a decision node (c.f. first column of Table III-(d)).

## C. Semantic Transformations: Fault Propagation Chains to Fault Tree

The probability definitions proposed for the analysis of FPCs are based on a theoretical concept where the failed execution of an action (system function), i.e. $a_i$, can be somehow observed experimentally. However, in reality, a failed execution of a system function may only be observed when the function is implemented, for example on hardware. In terms of system behavior and structure modeling, this is often referred to as allocation of system functions to system components. For example, if a system function, $A_i$, is allocated to two system components, then each individual failure rate of the two components will contribute to the calculation of the probability of functional failure, $P(a_i)$. As such, a different allocation strategy will likely result in different Fault Tree structure when physical components are considered. As this paper is only concerned with system behavioral faults, we push the allocation mechanism into the next-level of detail in the decomposition process of FTA, and this will be a topic of a future work.

Taking the FPC in Fig. 3(a) as an example, with a simplified situation where $a_i$ is the first fault event in the entire chain, i.e. $P(a_{i-}) = 0$, the FPC can be transformed into a Fault Tree that consists of an OR-gate as depicted in Fig. 4(a). This is because that based on (14), $P(a_{j+})$ can be considered as a union of $P(a_i)$ and $P(a_j)$, where $a_i$ and $a_j$ are basic events, and $a_{j+}$ is an intermediate event that can further contributes to the probability calculation of fault events after $a_j$. If there is no fault even after $a_j$, then the intermediate event $a_{j+}$ can be replaced by the top event, "system failure", $a_s$.

For the FPC in Fig. 3(b), following the transformation developed above, one can also translate the set of equations in (14) exactly into Fault Tree probability calculations in the following way:

$$\begin{cases} P(a_{i+}) = P(a_0 \text{ OR } a_i) \\ P(a_{j+}) = P(a_0 \text{ OR } a_j) \\ P(a_{k-}) = P(a_{i+} \text{ OR } a_{j+}) \\ P(a_{k+}) = P(a_{k-} \text{ OR } a_k) \end{cases}, \quad (19)$$

with

$$\begin{aligned} P(a_i \text{ OR } a_j) &= P(a_i \cup a_j) \\ &= P(a_i) + P(a_j) - P(a_i \cap a_j), \end{aligned} \quad (20)$$

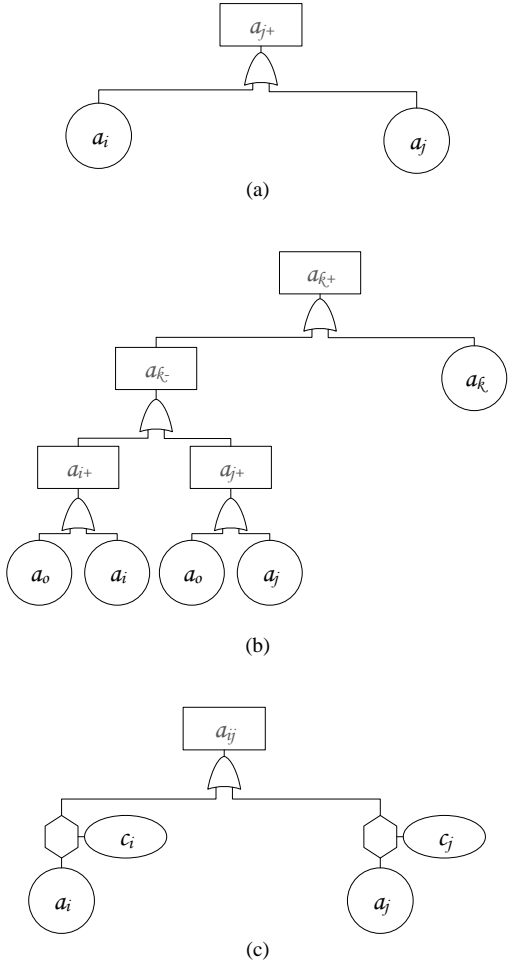

Fig. 4. Examples of Fault Trees transformed from various Fault Propagation Chains where (a) is transformed from FPC in Fig. 3(a); (b) is transformed from FPC in Fig. 3(b); and (c) is derived from an expanded contracted fault event.

and the intersection

$$P(a_i \cap a_j) = P(a_i)P(a_j|a_i) = P(a_j)P(a_i|a_j), \quad (21)$$

if $a_i$ and $a_j$ are dependent events. And the intersection reduces to

$$P(a_i \cap a_j) = P(a_i)P(a_j), \quad (22)$$

if $a_i$ and $a_j$ are independent.

The new set of equations in (19) can be immediately depicted graphically in a Fault Tree, as shown in Fig. 4(b), by defining $a_0$, $a_i$, $a_j$, and $a_k$ as basic events (independent of each other), and $a_{i+}$, $a_{j+}$, $a_{k-}$, and $a_{k+}$ as intermediate events.

For a contracted fault event, at the contraction level, it can be treated similar to a normal fault event. Hence, the mapping to Fault Tree is straightforward. For the fault events embedded in the contraction, one can translate (18) into

$$P(a_{i,j}) = P((a_i \cap c_i) \cup (a_j \cap c_j)). \quad (23)$$

The Fault Tree corresponding to this equation is depicted in Fig. 4(c). Because $c_i$ and $c_j$ are not fault events, but conditional statements, instead of using the normal AND-gates, inhibit-gates are used. Moreover, expanding the expression on the right hand side of the above equation, one would expect an intersection term that reads, $(a_i \cap c_i) \cap (a_j \cap c_j)$. However, this term yields zero as $c_i$ and $c_j$ are mutually exclusive events.

### D. Nested Control Flows

In this section, we demonstrate that the transformation method is not limited to simple connections of elementary control flows, but is also applicable to complex control flows that involve the nesting of elementary control flows. We apply the transformation method to the nested control flow shown in Fig. 5(a) to obtain a semantically equivalent fault tree. It is worth noting that in this nested control flow, Action, $A_j$ does not proceed to the Action, $A_m$.

Using the elementary control flows as the basis, the structure of the nested control flow can be logically expressed by the following expressions,

$$\begin{cases} (p_i \vee p_j) \to p_0 \\ p_n \to (p_i \vee p_j) \\ (p_m \vee p_n) \to p_i \\ p_f \to (p_m \vee p_n) \end{cases}, \quad (24)$$

Applying contraposition to the expressions in (24), we arrive at the following elementary units,

$$\begin{cases} (a_0 \to a_i) \wedge (a_0 \to a_j) = a_0 \to (a_i \wedge a_j) \\ (a_i \to a_n) \vee (a_j \to a_n) = (a_i \wedge a_j) \to a_n \\ (a_i \to a_m) \wedge (a_i \to a_n) = a_i \to (a_m \wedge a_n) \\ (a_m \to a_f) \vee (a_n \to a_f) = (a_m \wedge a_n) \to a_f \end{cases}, \quad (25a)$$

or alternatively, using the concept of contracted fault event, we have,

$$\begin{cases} a_0 \to a_{i,j} \\ a_{i,j} \to a_n \\ a_i \to a_{m,n} \\ a_{m,n} \to a_f \end{cases}, \quad (25b)$$

Although the conjunction of the above relations,

$$(a_0 \to a_{i,j}) \wedge (a_{i,j} \to a_n) \wedge (a_i \to a_{m,n}) \wedge (a_{m,n} \to a_f), \quad (26)$$

is in CMF, this is not a chain as described in (7) because it breaks at $\cdots a_n) \wedge (a_i \cdots$. To resolve this issue, we introduce only one contracted fault event, $a_{m,n}$, such that the FPC can be expressed by a disjunction of the two relations,

$$\begin{cases} (a_0 \to a_i) \wedge (a_i \to a_{m,n}) \wedge (a_{m,n} \to a_f) \\ (a_0 \to a_j) \wedge (a_j \to a_n) \wedge (a_n \to a_f) \end{cases}, \quad (27)$$

in which these two chains are two mutually exlusive paths. We emphasize that as a FPC, (27) is not in CMF due to having a disjunction in between the two relations.

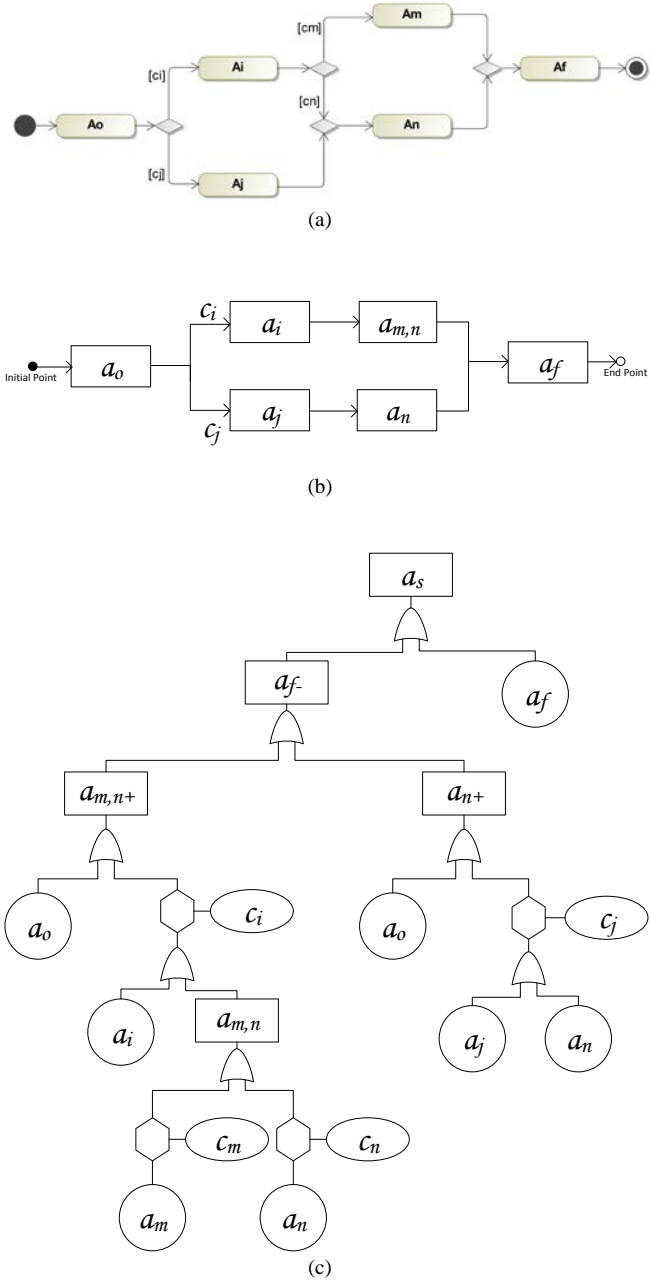

Fig. 5. A nested control flow example with (a) the nested control flow modeled in UML Activity; (b) a derived FPC as described in (27) for the nested control flow; and (c) The transformed Fault Tree for the nested control flow.

Continuing with (27), we depict the FPC in Fig. 5(b). In addition to the bifurcation, we also indicate the necessary condition for each of the exclusive chains resulted from the disjunction of the two relations in (27). Based on the transformation method deveoepd in the previous subsection, we transform the FPC into a Fault Tree that is depicted in Fig. 5(c), in which the detailed probability analysis for the FPC in Fig. 5(b) is provided in Appnedix I. The minimal cut sets of this Fault Tree are: $\{a_o\}$, $\{a_i, c_i\}$, $\{a_j, c_j\}$, $\{c_i, c_m, a_m\}$, $\{c_i, c_n, a_n\}$, $\{a_j, c_n\}$, and $\{a_f\}$.

The nested control flow example demonstrates that our method can be applicable to complex control flows in which the five elementary control flows as shown Table III are used as building blocks. However, to do an exhaustive investigation of the application of our method to all possible patterns is beyond the scope of this paper.

### E. The Overarching Metamodel

This section abstracts the developed rules of mapping an Activity model to a FPC and rules of transforming the FPC to a Fault Tree into an Activity model-Fault Propagation Chain-Fault Tree (AM-FPC-FT) overarching metamodel as depicted in Fig. 6.

The overarching metamodel is composed of 12 metaclasses that are differentiated by a white-grey-dark scale (white, yellow and blue for the colored version). On the left, three metaclasses (shown in grey), ControlFlow, Action, and ValueSpecification (for guard), are inherited from the RAM. In the middle, four metaclasses, as show in white, are introduced for the metamodeling of FPCs. In particular, the elementary structure of a FPC is in the form of fault events connected by directed edges (material implications). Therefore, similar to ControlFlow and Action in the RAM, two metaclasses, FaultEvent and MaterialImplication are introduced to reflect this structure. In addition, FaultEvent in the FPC metamodel is further specialized into two types: SingleFault and ContractFault. On the right, five metaclasses (shown in dark), ORGate, InhibitGate, OutputEvent, BasicEvent, and ConditionalEvent, are inherited from the FTM. The metaclasses, where relevant, inherit the relations established in their domain metamodel. These include the relations between the three metaclasses inherited from the RAM. The metamodel is developed and depicted in a way such that it also reflects, from left to right, the transformation method in which an Activity model is mapped into a FPC and then transformed into a Fault Tree.

The mapping from an Activity model to a FPC is captured by a stereotype, «contrapositive», which is introduced to refer to the Semantic Mapping Rules shown in Table III. As these mappings are structure-based rather than component-based, instead of associating individual elements, we connect the ControlFlow – Action structure to the MaterialImplication – FaultEvent structure (circled in dash lines) via the «contrapositive» relation. As different elementary control flows would give different fault propagation structure, an additional note is attached to the «contrapositive» relation to indicate the detailed rules as captured in Table III.

Moving on with the transformation from FPCs to Fault Trees, we first define an «equivalence» stereotype to illustrate a relation in which an entity is mapped exactly onto another entity without the need of modifications. In particular, we have three transformations that satisfy an «equivalence» relation: (i) Single Fault is mapped exactly onto Basic Event; (ii) ContractFault is mapped exactly onto an OutputEvent (as per intermediate event); and (iii) Guard information as captured by ValueSpecification is mapped exactly onto a ConditionalEvent. All of the above exact mappings are apparent from Fig. 3 to Fig. 4. In addition to the «equivalence» stereotype, we also introduce an «expansion» stereotype to capture the expansion of a contracted fault event into basic events as seen in Fig. 4(c) and Fig. 5(c). Since, a contracted fault event contains at least two exclusive paths, the multiplicity for the BasicEvent on the «expansion» relation is defined as $2\cdots *$. Additional notes are given to each of these stereotyped relations to provide detail explanations on the specific transformation mechanism.

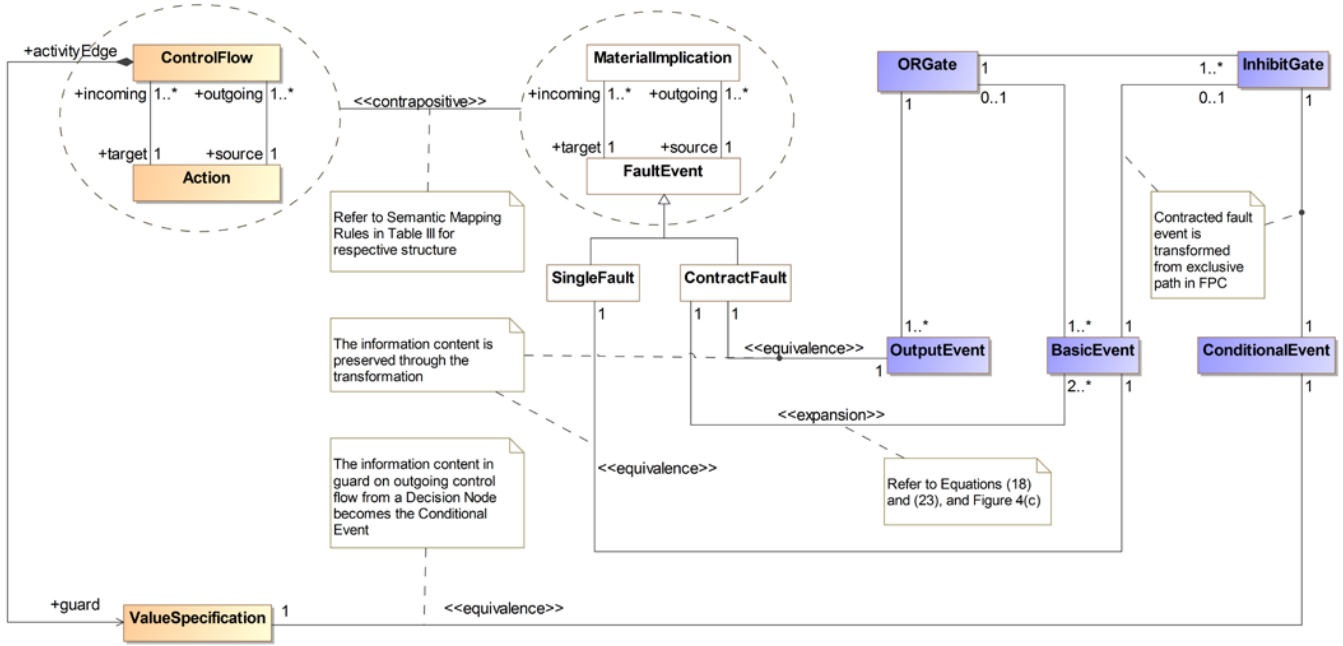

Fig.6. Activity model-Fault Propagation Chain-Fault Tree Overarching Metamodel.

As seen in the resultant Fault Trees, the transformations further impose a specific set of relationships between Fault Tree metaclasses. These are captured in the AM-FPC-FT metamodel. Firstly, we identify that an OR-gate can be connected directly to one or more inhibit-gates as seen in Fig. 3(c) and Fig. 4(c) due to the expansion of the contracted fault event. This is captured by an association between the metaclasses ORGate and InhibitGate with multiplicity 1 and $1\cdots *$ on each side. Secondly, depending whether a basic event is transformed from an Action that is within an exclusive path, i.e. proceed from satisfying its guard condition, or not, the basic event is either directly connected to an OR-gate or an inhibit-gate. Again, corresponding multiplicities are defined and additional notes are provided on the associations to provide detail explanation.

## VI. TRAFFIC MANAGEMENT SYSTEM OF SYSTEMS CASE STUDY

In this section, we evaluate our developed transformation method and the AM-FPC-FT overarching metamodel through an application of it to the Ramp Meter System (RMS) studied in [42],[43].

### A. The System and the Activity Model

The RMS is a constituent system of a Traffic Management System of Systems (TMSoS). The RMS is situated on the access ramp used to access inter-urban highways. The RMS employs two-phase (red and green) traffic lights to control the rate at which vehicles join the highway. The RMS has access to data about traffic in its own immediate vicinity as opposed to the Traffic Control Center (TCC) which has access to region wide traffic data. The RMS can operate in three different modes to control the traffic light. These modes are: (i) Fixed-time Mode with fixed length of red/green phases; (ii) Responsive Mode in which a varying phase is adopted to respond to logical traffic; and (iii) Collaborative Mode in which the control strategy is provided by the TCC.

We have modified the Concept of Operation (CoO) of the RMS slightly to allow both concurrent control flows and alternative control flows to be presented in the Activity model, and to avoid control loops that are not within the scope of this paper. The RMS begins each traffic light control cycle with collecting local data (vehicle flow rate). Then, the RMS starts to analyze the data, and concurrently, send the data to the TCC. Based on the data received, the TCC will decide whether to instruct the RMS to operate in Collaborative Mode. The analysis result by the RMS and the instruction from the TCC together form the basis on which operational mode, i.e. Fixed-time Mode, Responsive Mode, and Collaborative Mode, to be selected. The decision logic for mode selection follows that TCC Instruction has higher priority over RMS analysis result (note that this logic will not be modeled explicitly as they are treated as a detailed design at a lower level). The RMS will then implement the mode being selected. The successful execution of a control cycle will lead to the traffic light to operate in a pre-defined mode. The execution of the next control cycle may or may not change the mode being previously operated.

To model the behavior of the RMS, a Use Case Diagram, as depicted in Fig. 7, is constructed to capture all of the system functions given in the CoO. Then, the functions are ordered into control flows based on the details of the CoO. The corresponding Activity model is presented in Fig. 8. In the

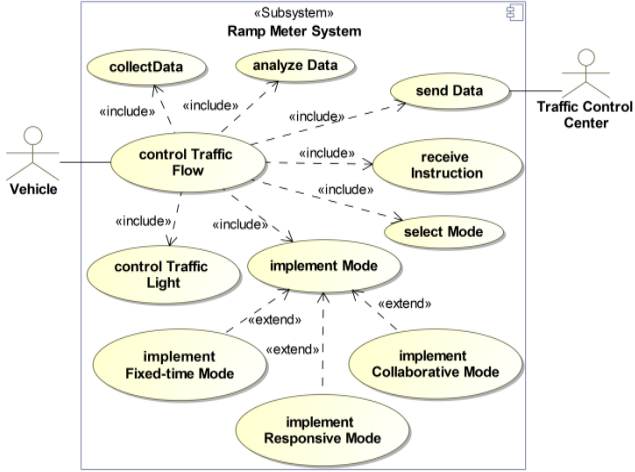

Fig. 7. The Use Case model of the Ramp Meter System.

Activity model, a system function (action) is modeled by a verb-noun phrase with the first letter of the nouns capitalized. In addition, a proposition denoted by $p_i$ is used to model the completion of the execution of the corresponding system action, $A_i$. The RMS control flow starts with a "collect Data" ($A_1$) action and flows into pair of fork and join nodes. In between the fork and the join node, two concurrent series of actions are modeled. One series on the left has a single action, "analyze Data" ($A_2$); and the other series on the right starts with a "send Data" ($A_3$) action that leads to an external function, "send Instruction" ($A_4$), owned by the TCC. And then, the series flows back to the RMS where RMS shall "receive Instruction" ($A_5$). Immediately after the join node, the control flow continues into a "select Mode" ($A_6$), where the logic of mode selection shall be modeled at the next level of detail. Once a decision is made, the control flow continues into a pair of decision and merge nodes in which three alternative paths are modeled. Each path involves an action of a designated mode implementation, i.e. "implement Fixed-time Mode" ($A_7$) with "Fixed-time Model selected" ($c_7$), "implement Responsive Mode" ($A_8$) with "Responsive Model selected" ($c_8$), and "implement Collaborative Mode" ($A_9$) with "Collaborative Model selected" ($c_9$). With only one out of the three modes being implemented at any one time, the RMS then completes the control cycle with "control Traffic Light" ($A_{10}$) according to the operation mode implemented.

### B. RMS Fault Trees

In the original study of the RMS in [42], five faults of the RMS have been identified that could lead to non-optimal traffic flow. The faults are:
1. Lights stuck on green or no lights at all,
2. RMS fails to adopt Collaborative Mode when instructed,
3. Lights stuck on red,
4. RMS fails to exit Collaborative Mode when instructed,
5. RMS calculates an incorrect rate for vehicles to be admitted.

These five faults can be naturally organized into a conventional Fault Tree, as depicted in Fig. 9, with "Non-optimal traffic flow" being the top event. Then, the Fault

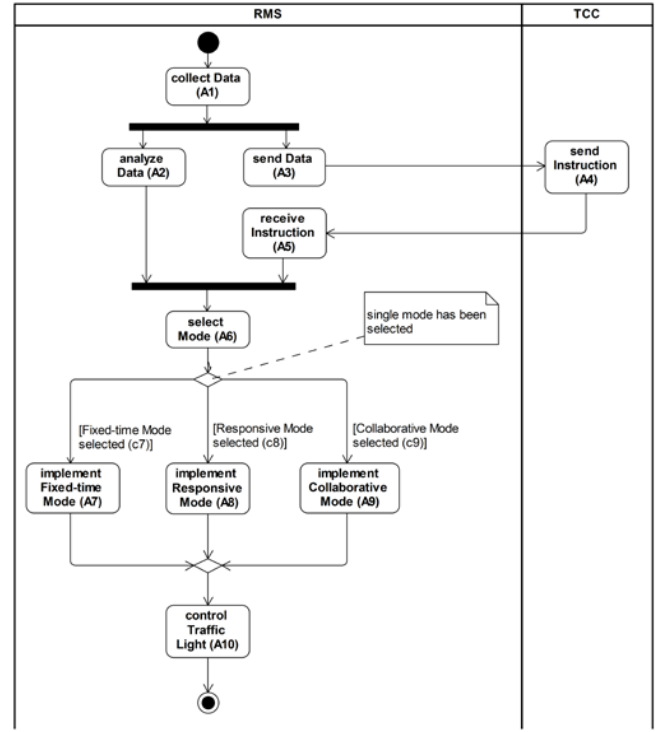

Fig. 8. The Activity model of the Ramp Meter System.

Tree separates faults into component-based (Faulty traffic light) and functional-based (Operational faults) through a top-down deduction. In the rest of this section, we will apply the transformation method to the Activity model in Fig. 8 and demonstrate how the generated Fault Tree can provide useful information in the identification of behavioral faults as well as inferring a fault structure. The fault structure is then used to analyze the original design model.

For convenience, we define a set of fault events as the negation of the propositions, $a_i \stackrel{\text{def}}{=} \neg p_i$. Based on the semantic mapping rules (a), (b) and (c), the initial fault event $a_1$ bifurcates into two chains, $(a_1 \rightarrow a_2) \wedge (a_1 \rightarrow a_3)$, with one chain has a single fault event, $a_2$, and the other chain has three fault events connected in a series, $(a_3 \rightarrow a_4) \wedge (a_4 \rightarrow a_5)$. The two chains then merge into a single chain that leads to the fault event, $a_6$ with $(a_2 \rightarrow a_6) \wedge (a_5 \rightarrow a_6)$. The merged chain then continues into a contracted event $a_{7,8,9} = a_7 \wedge a_8 \wedge a_9$. The contracted fault event, $a_{7,8,9}$, further propagates to the last fault event, $a_{10}$, through the unitary implication $a_{7,8,9} \rightarrow a_{10}$. Grouping all of the unitary implications together, a complete FPC is generated and depicted graphically in Fig. 10. An initial point and end point is added to the chain to show where the fault propagation starts and ends. These points map to the initial node and end node in the original Activity model.

Next, we transform the FPC to a Fault Tree using the established transformation method. The detailed physical architecture, i.e. allocation of functions to physical component will not be considered. Without going through the detailed mathematical derivation, one can obtain the Fault Tree by: firstly, elaborating the relations in (19) for the bifurcation and convergence of the two parallel chains $(a_1 \rightarrow a_2) \wedge (a_2 \rightarrow a_6)$ and $(a_1 \rightarrow a_3) \wedge (a_3 \rightarrow a_4) \wedge (a_4 \rightarrow a_5) \wedge (a_4 \rightarrow a_5)$; and then elaborating the relation in (23) for the contracted event, $a_{7,8,9}$ for three exclusive chains; and finally using the relations in (19) again to integrate everything to obtain the final RMS Fault Tree as depicted in Fig. 11. The top event of the RMS Fault Tree, system failure, $a_s$, is defined as the system failing to complete the intended system behavior as modeled in the UML Activity.

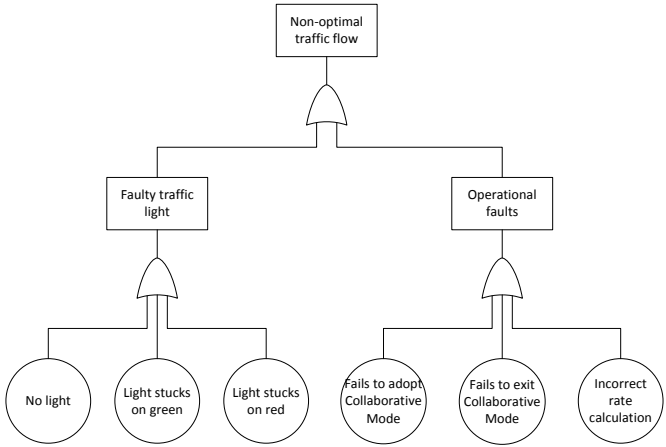

Fig. 9. A RMS Fault Tree constructed based on the faults identified in [43].

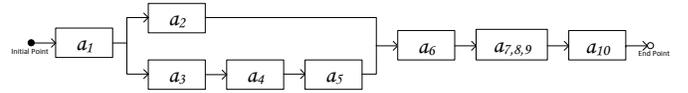

Fig. 10. The Fault Propagation Chain transformed from the RMS Activity model.

### C. Qualitative Analysis of the Transformed Fault Tree

Without a designated probability for each of the basic events, it is difficult to provide a meaningful quantitative analysis. Nonetheless, based on the structure of the RMS Fault Tree, we provide the following qualitative analyses:

Firstly, every system function, as modeled by actions, becomes a basic (fault) event in the final RMS Fault Tree after the transformations. The minimal cut sets are: $\{a_1\}$, $\{a_2\}$, $\{a_3\}$, $\{a_4\}$, $\{a_5\}$, $\{a_6\}$, $\{a_7, c_7\}$, $\{a_8, c_8\}$, $\{a_9, c_9\}$, and $\{a_{10}\}$. The single event minimal cut sets align with the formal interpretation of the Activity model based on the UML specification where failed execution of an action stops the control flow. In addition, fault events $a_7$, $a_8$ and $a_9$ do not individually lead to system failure unless their corresponding condition are met. This also aligns with the Activity model in which the three alternative paths are exclusive, i.e. only one mode is adopted at a time. Hence, despite the detailed structure of the Fault Tree, it is obvious that ensuring a high reliability for each of the designed system function is a straightforward way to minimize the probability of the occurrence of the top event.

Secondly, on the right-hand side of the RMS Fault Tree, the set of inhibit-gates used for the contracted fault event, $a_{7,8,9}$, implies that the actual contributions from $P(a_7)$, $P(a_8)$, and $P(a_9)$ to $P(a_{7,8,9})$ is suppressed subject to the statistical probabilities of their corresponding mode selection. For instance, if Responsive Mode is rarely selected for a particular local RMS, the contribution of $P(a_8)$ to the top event will be insignificant (note that this is not to say that the "implement Responsive Mode" function is insignificant). Reversely, ensuring a high reliability for the operational mode mostly selected can reduce the top event probability.

Lastly, on the left-hand side of the RMS Fault Tree, we observe that one of the basic event, $a_4$, can be considered as an external event. Tracing this fault event and its surrounding Fault Tree structure to the original functional design, we realize that if TCC fails to "send Instruction" to the RMS or the RMS fails to "receive Instruction" from the TCC, the system will not be able to "select Mode". Instead of letting the system stuck at this point, the designer may want the system to treat this situation as if no TCC instruction is received. As such, though the local RMS will not operate in an optimal mode, it can continue to operate. The designer then has various choices to model this mechanism. For instance, the modeling can be done through revising the current Activity model to include an additional decision point or through embedding this decision logic into the next level detail under the function, "select Mode".

Finally, we compare the transformed Fault Tree in Fig. 11 with the Fault Tree derived from [42] and depicted in Fig. 9. Although the top events in the two Fault Trees are different, we recognize that the top event, system failure, $a_s$, in fact, contributes to the top event, non-optimal traffic flow as in Fig. 9. This is because when system failure happens, vehicles entering the ramp would not be controlled under the desired strategy; hence leading to a non-optimal traffic flow. In addition, overlaps between the two Fault Trees are identified. 'Fails to adopt Collaborative Mode' as captured in the original Fault Tree is also captured in the transformed Fault Tree by the basic event, $a_9$. However, as suggested by the transformed Fault Tree, this fault event itself will not be a minimal cut set. In fact, it will only lead to system failure and eventually non-optimal flow when the system decides to adopt Collaborative Mode. Similarly, the fault event, 'Fails to exit Collaborative Mode' can be considered as an overlap to $a_7$ and $a_8$ collectively in which the system situationally decides to adopt one of the two other modes when it is originally in the Collaborative Mode. Again, an inhibit-gate is necessary in these two cases. From the above arguments, we conclude that the developed transformation can firstly identify additional basic events based on system behavior models; and secondly can provide meaningful revisions to a Fault Tree constructed by engineers. From the comparison, it is also clear that due to the scope of the transformation method, the transformed Fault Tree does not capture component failures and faults that relate to objects, e.g. incorrect rate calculation.

To summarize the case study evaluation, a Fault Tree has been successfully generated by applying our transformation method to the RMS Activity model. As the generated Fault Tree preserves structural knowledge of the original model, we are able to analyze qualitatively the Fault Tree to derive important system reliability information and provide design improvements. By comparing the transformed Fault Tree to the one derived based on common practice, the transformation can provide meaningful addition and revision.

*D. A Comparison against other Transformation Methods*

This subsection evaluates the method by conducting a qualitative comparison against established transformation methods. For this comparison, we categorize the established transformation methods into two sets. The first set concerns with transforming system models developed in standardized modeling languages such as UML [18],[19],[20],[21] and SysML [22],[23],[24] to Fault Trees; while the second set transforms information modeled in other techniques such as diagraphs [4],[5], component-based modeling [2],[3], knowledge-based approach [8], and architecture description language [44]. Our method is a transformation for system models, i.e. it transforms a UML Activity model into a Fault Tree. Furthermore, a comparison of the two sets of methods would require a comparison between the modeling techniques, which would be beyond the scope of this paper. Therefore, we will only compare our method against the first set of methods.

Our method is unique in two ways: firstly, it is the first attempt in transforming an Activity model with a focus on system behavior and fault propagation; secondly, it has a formal basis. In comparison, on the UML side, similar to our work, there are methods developed based on system model presented in a single type of diagram: Class Diagram in [18]; Use Cases in [19] and State Machine in [21]. Some other methods developed consider multiple diagrams: for instance, [20] has a wider focuses by utilizing UML Composite Structure, Sequence and Use Case Diagrams. On the SysML side, methods developed in [22] transform information captured in SysML Internal Block Diagrams; and the method developed in [24] utilizes both Block and Internal Block Diagrams. Although many of the above methods have considered system behaviors, none of them emphasizes the concept of fault propagation that is manifested from control flows as modeled in UML or SysML Activities.

In addition, to facilitate the transformation methods developed in most of these works, additional stereotypes are introduced to the standard UML and SysML models; whereas our method does not. As such, we argue that the advantage of our method is that it reflects exactly the reliability of system behavior architecture; while the limitation of the method is that it only concerns one type of fault, which is functional failure. Additionally, the formal basis of our method, which is lacking in the other established methods, could give engineers more confidence in the correctness of a generated Fault Tree generated by our method.

We anticipate that a more sophisticated comparison will be possible in our future work in which our current method will be elaborated to include the consideration of system physical architecture modeled, and the allocation of activities to UML Classes.

VII. CONCLUSION AND FUTURE WORK

This paper developed a transformation method based on propositional logic and probability theory to allow control flows modeled in UML Activities to be transformed into semantically equivalent Fault Trees. The developed method aligns with current industrial practices in early stage system assurance [9] and advances existing approaches in terms of accommodating system model availability [2],[3] and incorporated mathematical rigor [22],[24]. We introduced a new concept, FPC, as an intermediate step, to facilitate the transformation method. An AM-FPC-FT overarching metamodel was then developed to abstract the transformations among Activity models, FPC and Fault Trees. The formal basis of the transformation method together with the overarching metamodel facilitate automated Fault Tree generation. Furthermore, the formal basis of the transformation method guarantees the generated Fault Tree to be semantically correct. Our method reveals an important finding where the formal approach suggests that mappings should be based on the relational structure between entities and not just entity-to-entity.

To demonstrate and evaluate the applicability of our method, the developed transformation method was applied to a Ramp Meter System case study previously studied in [42] and [43]. A Fault Tree qualitative analysis was then carried out to examine the quality of the functional design. Based on the structural implications of the RMS Fault Tree, potential alterations were suggested.

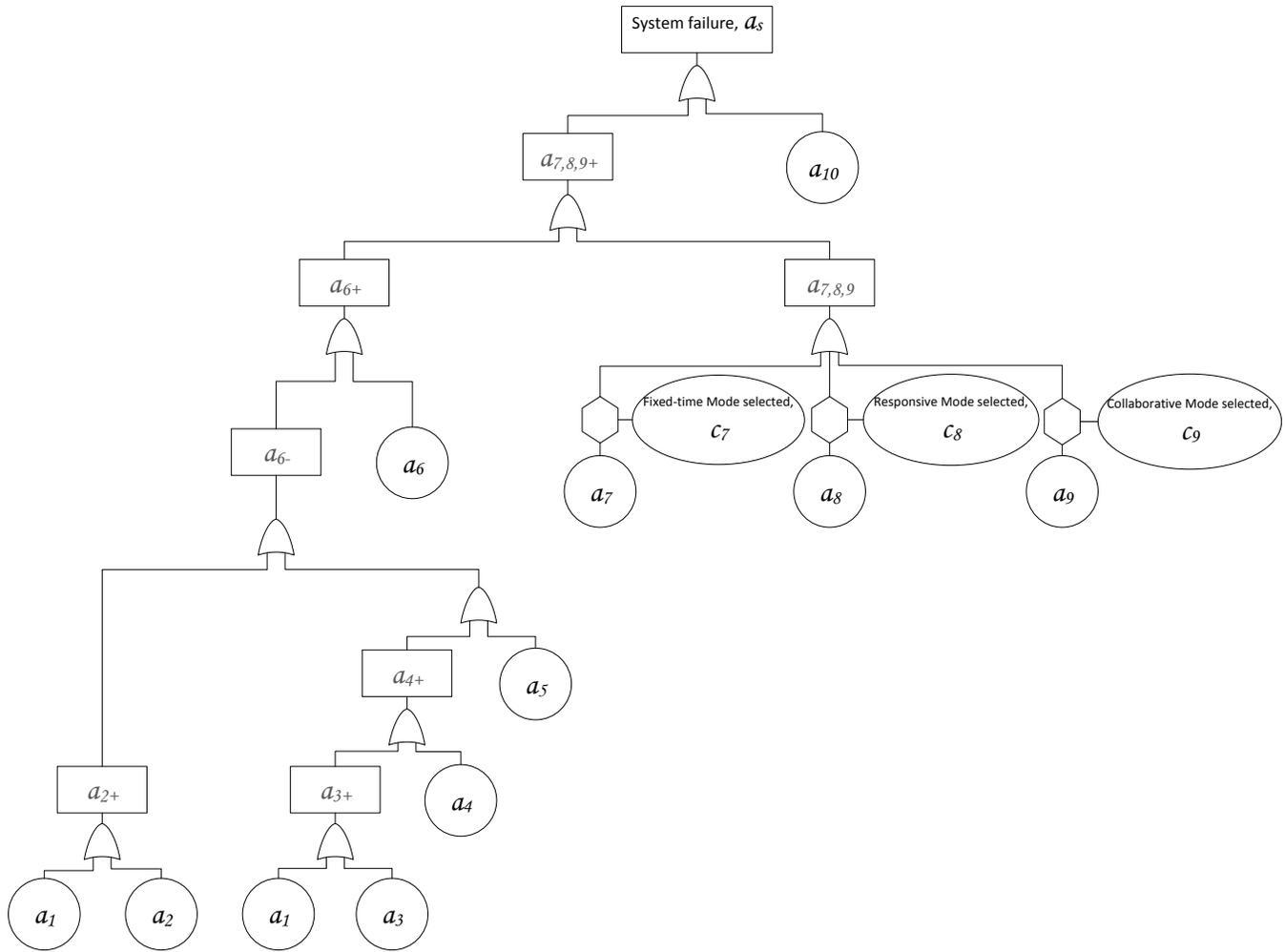

Fig. 11. The RMS Fault Tree transformed from the Fault Propagation Chain in Fig. 10.

Despite the successful application of the method to the case study, there are limitations of the approach presented in this work. For instance, loops are particularly useful in modeling control systems; but have been deliberately avoided in control flows as they do not naturally fit into the structure of a standard Fault Tree. Majdara and Wakabayashi in [2] also suggested that events that are encountered repeatedly can lead to an infinite loop structure in a Fault Tree. Therefore, it is essential to resolve this issue with the usage of UML Structured Activity Nodes, such as LoopNode in a future work.

Furthermore, because the current state of the approach is concerned with only propositional logic and control flows, the transformation method is not applicable to object flows and only considers one type of fault, i.e. a system function fails to complete execution. As such, to make our method more widely applicable, three pathways are envisioned to extend the current work. These are extensions of the transformation of control flows to: (i) object flows, (ii) State Machines, and (iii) functional allocations. The following discusses each pathway in more details. With these extensions, it should then be possible to evaluate the reliability of the transformed Fault Tree in great depth by comparing it to similar works.

Object flow, which was not included in the scope of our current work, is the other important aspect of UML Activities. Errors presented in an Object, e.g. incorrect data, that are passed between system components may lead to undesired system behaviors such as malfunctions in addition to the fault events concerned in the current work. Developing a formal representation of Objects and their associated faults can be challenging. It is anticipated that the use of Predicate Calculus to capture states of an Object can provide insights in the inclusion of object flows.

The modeling of system behavior is also concerned with the modeling of system states. In UML, these system states are modeled using State Machines. As the executions of system functions are closely related to the changing of system states, transformation between the two types of behavioral models could provide insights in elaborating our current transformations to further consume State Machines as system behavioral models. It would be also useful to utilize and elaborate other existing works such as [21].

Lastly, in practice, system functions are implemented in physical components. Therefore, it is essential to extend the current method to include aspects related to system components and functional allocations. UML uses Classes to model the system structure, system components and functional allocations. Hence, it is natural to elaborate the current work to develop transformation methods that map system structures and components modeled using Classes into semantically equivalent Fault Trees. This should further advance existing approaches detailed in [18], [22] and [24].

## APPENDIX I

To demonstrate that the FPC in Fig. 5(b) indeed maps into the Fault Tree depicted in Fig. 5(c), we perform a detailed probability analysis for this FPC. Again, we use the concept of running the system for $n$ times and observing the number of system failure.

By running the system for $n$ times, based on the derivation of (14), the expected number of system failure, $q_1$, upto the point at $a_{0+}$ can be determined as,

$$q_1 = nP(a_0), \tag{28}$$

The FPC then bifurcates into two chains with exlusive conditions after the fault event, $a_0$. The number of system success that goes through the top chain with condition, $c_i$, can be calculated as,

$$m = n(1 - P(a_0))P(c_i), \tag{29}$$

Then, we can calculate the additional number of system failure at $a_{i+}$ as,

$$q_2 = mP(a_i), \tag{30}$$

and the additional number of system failure at $a_{m,n+}$ as,

$$q_3 = m(1 - P(a_i))P(a_{m,n}), \tag{31}$$

where $P(a_{m,n})$ is the probability of the contracted fault event, $a_{m,n}$, and can be calculated based on the formula derived in (18). Using (28) to (31), we can calculate $P(a_{m,n+})$ by

$$P(a_{m,n+}) = \frac{q_1 + q_2 + q_3}{n}$$
$$= P(a_0) + (1 - P(a_0))P(c_i)\big(P(a_i) + (1 - P(a_i))P(a_{m,n})\big)$$
$$= P(a_0) + (1 - P(a_0))P(c_i)P(a_i \cup a_{m,n})$$
$$= P\Big(a_0 \cup \big(c_i \cap (a_i \cup a_{m,n})\big)\Big)$$
$$= P\Big(a_0 \cup \big(c_i \cap \big(a_i \cup ((a_m \cap c_m) \cup (a_n \cap c_n))\big)\big)\Big), \tag{31}$$

where the last step has used the derived result provided in (23) for a contracted fault event.

For the bottom chain, condition $c_j$ is required to proceed to $a_j$. Hence, similar to (29), the number of system success that goes through the bottom chain with condition, $c_j$, can be calculated as,

$$m' = n(1 - P(a_0))P(c_j), \tag{32}$$

Then, we can calculate the additional number of system failure at $a_{j+}$ as,

$$q_4 = m'P(a_j), \tag{33}$$

and the additional number of system failure at $a_{n+}$ as,

$$q_5 = m'\big(1 - P(a_j)\big)P(a_n), \tag{34}$$

Using (28), (32) to (34), we can calculate $P(a_{n+})$ by,

$$P(a_{n+}) = \frac{q_1 + q_4 + q_5}{n}$$
$$= P(a_0) + (1 - P(a_0))P(c_j)\big(P(a_j) + \big(1 - P(a_j)\big)P(a_n)\big)$$
$$= P(a_0) + (1 - P(a_0))P(c_j)P(a_i \cup a_j)$$
$$= P\Big(a_0 \cup \big(c_j \cap (a_i \cup a_j)\big)\Big). \tag{35}$$

For the convergence of the two chains, we use the third relation in (15) and in (19), to obtain $P(a_{f-})$, where

$$P(a_{f-}) = P(a_{m,n+}) + P(a_{n+}) - P(a_{m,n+})P(a_{n+}|a_{m,n+})$$
$$= P(a_{m,n+} \cup a_{n+}), \tag{36}$$

with $P(a_{m,n+})P(a_{n+}|a_{m,n+})$ containing the repeated events, $a_0$. Finally, we simply use the derived relation in (12) to obtain the top event probability, $P(s)$, as

$$P(s) = P(a_{f+}) = P(a_{f-}) + P(a_f) - P(a_{f-})P(a_f)$$
$$= P(a_{f-} \cup a_f). \tag{37}$$

In the above transformations, we have used basic events, $a_0$, $a_i$, $a_j$, $a_m$, $a_n$, and $a_k$; conditional events, $c_i$, $c_j$, $c_m$, and $c_n$; and specifically defined intermediate events $a_{m,n+}$, $a_{n+}$, and $a_{f-}$. Representing relations in (31), (35), (36), and (37) all together gives the Fault Tree depicted in Fig. 5(c).

ACKNOWLEDGMENT

The authors thank Mr. Mole Li for his suggestions on the metamodel construction.

**Charles E. Dickerson** (M'06–SM'17) received the Ph.D. degree from Purdue University, West Lafayette, IN, USA, in 1980.

He is Chair of Systems Engineering at Loughborough University and a Principal Investigator of complex vehicle system analysis in the Programme for Simulation Innovation. His research and design experience includes MIT Lincoln Laboratory, the Lockheed Skunkworks and Northrop Advanced Systems.

Prof. Dickerson is Chair of the Mathematical Formalisms Group at the OMG and the Assistant Director for Analytic Enablers at INCOSE.

**Rosmira Roslan** received the M.E. degree in Manufacturing Systems Engineering from Universiti Putra Malaysia, Selangor, Malaysia, in 2012.

She is currently pursuing the Ph.D. degree in systems engineering at Loughborough University, Loughborough, United Kingdom. From 2010 to 2015, she was a Quality Engineer with the Composites Technology Research Malaysia Systems Integration.

Ms. Roslan is a registered graduate engineer of Board of Engineers Malaysia since 2009.

**Siyuan Ji** (M'16) received the Ph.D. degree in physics from University of Nottingham, Nottingham, United Kingdom, in 2015.

He is currently a Research Associate at Wolfson School of Mechanical, Electrical, and Manufacturing Engineering of Loughborough University, United Kingdom. His research focuses on mathematical-based systems engineering methodologies and design algorithms for complex systems and system of systems.

Dr. Ji is a member of the IET and a member of the OMG.